\documentstyle{cupconf}

\input psfig

\ifoldfss
\else
  \ifnfssone
    \newmathalphabet{\mathit}
      \addtoversion{normal}{\mathit}{cmr}{m}{it}
      \addtoversion{bold}{\mathit}{cmr}{bx}{it}
    \newmathalphabet{\mathcal}
      \addtoversion{normal}{\mathcal}{cmsy}{m}{n}
    \else
    \ifnfsstwo
    \fi
  \fi
\fi

\def\hexnumber#1{\ifcase#1 0\or1\or2\or3\or4\or5\or6\or7\or8\or9\or
 A\or B\or C\or D\or E\or F\fi }

\makeatletter
\ifx\CUP@mtlplain@loaded\undefined
\else
\fi
\makeatother
 \makeatletter
 \ifx\CUP@mtlplain@loaded\undefined
   \font\tenbmi=cmmib10 at 10pt
   \font\sevenbmi=cmmib10 at 7pt
   \font\fivebmi=cmmib10 at 5pt

   \newfam\bmifam
   \textfont\bmifam=\tenbmi
   \scriptfont\bmifam=\sevenbmi
   \scriptscriptfont\bmifam=\fivebmi
   
 \fi
 \makeatother
\ifnfsstwo

\fi
\ifnfssone

\fi
\ifoldfss

\fi

\mathchardef\varLambda="0103

\newcommand{\kms}{$km~s^{-1}$}
\newcommand{\eg}{{\it e.g., }} 
\newcommand{\ie}{{\it i.e.~}} 
 
\newcommand{\etal}{{\it et al.}} 
\newcommand{\msun}{$M_\odot$ } 
\newcommand{\mloss}{$M_\odot~yr^{-1}$ } 
\newcommand{\apj}{{\it Astrophys. J. }} 
\newcommand{\aap}{{\it Astron.\& Astrophys. }}
\newcommand{\aaps}{{\it Astron. \& Astrophys. Suppl. }}
 
\newcommand{\apjs}{{\it Astrophys. J. Suppl. Ser. }} 
 
\newcommand{\mnras}{{\it Mon. Not. Roy. Astron. Soc. }}

\makeatletter
\ifx\CUP@mtlplain@loaded\undefined
\else
\fi
\makeatother
\makeatletter
\ifx\CUP@mtlplain@loaded\undefined
  \font\tenbms=cmbsy10
  \font\sevenbms=cmbsy10 at 7pt
  \font\fivebms=cmbsy10 at 5pt
  \newfam\bmsfam
  \textfont\bmsfam=\tenbms
  \scriptfont\bmsfam=\sevenbms
  \scriptscriptfont\bmsfam=\fivebms

  \edef\bsy@{\hexnumber\bmsfam}
  \mathchardef\bnabla="0\bsy@72
\fi
\makeatother
\def\eg{{e.g.\ }}

\def\etal{\mbox{\it et al.}}

\begin{document}
\ifnfssone
\else
  \ifnfsstwo
  \else
    \ifoldfss
      \let\mathcal\cal
      \let\mathrm\rm
      \let\mathsf\sf
    \fi
  \fi
\fi

  \title[Pre-SN Evolution of Massive Stars]{Pre-Supernova Evolution of
	Massive Stars} 

  \author[N. Panagia \& G. Bono]{%
  N\ls I\ls N\ls O\ns 
  P\ls A\ls N\ls A\ls G\ls I\ls A$^{1, 2}$,\ns 
  \and\ns
  G\ls I\ls U\ls S\ls E\ls P\ls P\ls E\ns 
  B\ls O\ls N\ls O$^3$}
  \affiliation{$^1$Space Telescope Science Institute, 3700 San Martin 
	Drive, Baltimore, MD 21218, USA. \\[\affilskip]
    	$^2$On assignment from the Space Science Department of ESA.\\[\affilskip]
	$^3$Osservatorio Astronomico di Roma, Via Frascati 33, Monte Porzio 
	Catone, Italy.}

    
  \maketitle

\begin{abstract}
We present the preliminary results of a detailed theoretical
investigation  on the hydrodynamical properties of Red Supergiant
(RSG)  stars at solar chemical composition and for stellar masses
ranging from  10 to 20 $M_\odot$. We find that the main parameter
governing their  hydrodynamical behaviour is the effective temperature,
and indeed when moving from higher to lower effective temperatures the
models show an increase in  the dynamical perturbations. Also, we find
that  RSGs are pulsationally unstable for a substantial portion of
their lifetimes. These dynamical instabilities play a key role in
driving mass loss, thus inducing high mass loss rates (up to almost
10$^{-3}$ \mloss) and considerable variations of the mass loss activity
over timescales of the order of  10$^4$ years. Our results are able to
account for the variable mmass loss rates as implied by radio
observations of type II supernovae, and we anticipate that comparisons
of model predictions with observed circumstellar phenomena  around SNII
will provide valuable diagnostics about their progenitors and their
evolutionary histories.

\end{abstract}

\firstsection 

\section{Introduction}

More than 20 years of radio observations of supernovae (SNe) have
provided a wealth of evidence for the presence of substantial amounts
of  circumstellar material (CSM) surrounding the progenitors of SNe of
type II and Ib/c (see Weiler \etal, this Conference, and references
therein). Also,  the radio measurements indicate that $(a)$ the CMS
density falls off like $r^{-2}$, suggesting a constant velocity, steady
wind, and that $(b)$ the density is so high as to require a ratio of
the mass loss rate, $\dot{M}$, to the wind velocity, $w$, to be  higher
than $\dot{M}/w\sim 10^{-7}$ \mloss \kms. These requirements are best
satisfied by red supergiants (RSG), with original masses in the range
8-30 \msun,  that indeed are the putative progenitors of SNII. Note
that in the case of SNe Ib/c, the stellar progenitor cannot provide
such  a dense CSM directly and that a wind from a binary companion must
be invoked to explain the observations (Panagia and Laidler 1988, Boffi
and Panagia 1996, 2000). 

This scenario is able to account for the basic properties of all radio
SNe. However, the evolution of SN 1993J indicated that the progenitor
mass loss rate had  declined by almost a factor of 10 in the last few
thousand years before explosion (Van Dyk \etal~ 1994).  In addition,
there are SNe, such as SN 1979C (Montes \etal~ 2000), SN 1980K (Montes
\etal~ 1998), and SN 1988Z (Lacey \etal~ 2000), that have displayed 
relatively sudden changes in their radio emission evolution about 10
years after explosion, which also cannot be explained in term of a
constant mass loss rate. Since a SN shock front, where the radio
emission originates, is moving at about 10,000~\kms~ and a RSG wind is
typically expanding at 10~\kms, a sudden change in the CSM density
about ten years after explosion implies a relatively quick change of
the RSG mass loss rate about 10,000 years before it underwent the SN
explosion.  These findings are summarized in Figure 1 that, for several
well studied RSNe,  displays the mass loss rate implied by radio
observations as a function of the look-back time, calculated simply as
the actual  time since explosion multiplied by a factor of  1000, which
is  the ratio of the SN shock velocity to the RSG wind velocity. 

Additional evidence for enhanced mass loss from SNII progenitors over
time intervals of several thousand years is provided also by the
detection of relatively narrow emission lines with typical widths of
several 100~\kms~ in the spectra of a number of SNII (\eg SN 1978K:
Ryder \etal~ 1993, Chugai, Danziger \& Della Valle 1995, Chu
\etal~1999; SN 19997ab: Salamanca \etal~1998; SN 1996L: Benetti \etal~
1999), that indicate the presence of dense circumstellar shells ejected
by the SN progenitors in addition to a more diffuse, steady wind
activity.

 We note that a time of about 10,000 years is a sizeable fraction of
the time spent by a massive star in the RSG phases and implies a kind
of variability which is not predicted by standard stellar evolution. 
In particular, a time scale of $\sim 10^4$ years is considerably
shorter than the H and He burning phases but is much longer than any of
the successive nuclear burning phases that a massive star goes through
before core collapse (\eg Chieffi et al. 1999). Therefore, some other
phenomenon is to be sought to  properly account for the observations.

\begin{figure}
\centerline{\psfig{figure=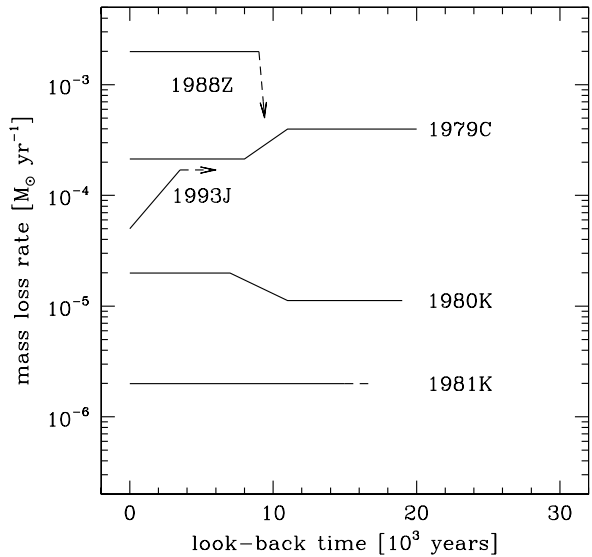,width=8.5cm}}
\caption{Mass loss rates as a function of look-back time as measured
for a number radio supernovae (adapted from Weiler \etal~2000; 
schematic)}
\label{radiomloss}
\end{figure}

\begin{figure}
\centerline{\psfig{figure=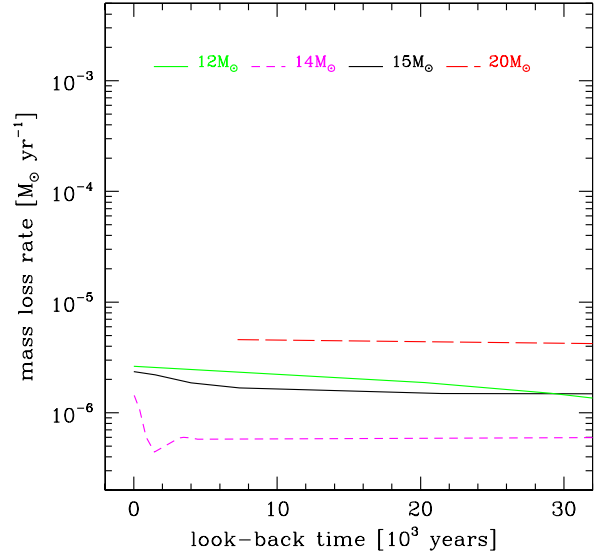,width=8.5cm}} 
\caption{Mass loss rates predicted from canonical stellar evolution theory
by using Reimers' formula.}
\label{doesntwork}
\end{figure}

\begin{figure}
\centerline{
\psfig{figure=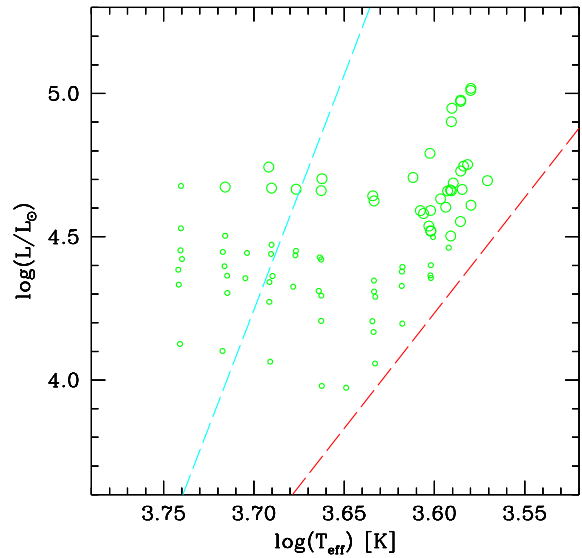,width=8.5cm}} 
\caption{Predicted mass loss rates as a function of the position in the
HR diagram adopting Reimers' formula. The size of the circles are
proportional to the logarithm of the mass loss rate: the highest values
are several 10$^{-6}$ \mloss~ and the lowest ones about 10$^{-7}$ \mloss. The
dashed lines represent the analytical relations for the fundamental
blue and red edges of the Cepheid instability strip defined by Bono
\etal~ (2000a).}
\label{Reimers_mloss} 
\end{figure}

Another problem which needs to be addressed is the actual rate of mass
loss for red supergiants. The observational evidence is that mass loss
rates in the range 10$^{-6}$--10$^{-4}$ \mloss are commonly found in
RSG, with a relatively steep increase in mass loss activity for the
coolest stars (\eg  Reid, Tinney \& Mould 1990, Feast 1991). On the
other hand, there is no statisfactory theory to predict mass loss rates
in these phases of stellar evolution, and current parametrizations fall
short from describing the phenomenon in detail. For example, let us
consider the classical formula by Reimers (1975),

$$log(\dot{M})~ =~ -12.6~ +~ log(\frac{LR}{GM})~ + ~log(\eta)$$

which can be rewritten as:

$$\dot{M}~ \propto ~\eta ~\frac{L^{1.5}}{M~T_{eff}^2} $$

\vskip .1in

\noindent This formula was devised to dimensionally account for the
mass loss from low-mass red giants, but has also been widely adopted
for evolutionary track calculations. We see that  the predicted mass
loss  rate varies rather slowly when a star is moving from the blue to
the red region (i.e. during H-shell burning  and/or He-core burning) of
the  HR diagram, the main functional  dependence being a 1.5 power of
the luminosity. The corresponding mass loss rates, computed using the
evolutionary tracks by Bono \etal~ (2000b) for stars in the mass
range 10--20\msun, are shown in Figures 2 and 3.  It is apparent that
not only the rates are not as high as suggested by spectroscopic
observations of RSGs  (this aspect alone could easily be ''fixed" by
increasing the efficiency factor $\eta$) but, more importantly, are
very slowly varying with time and, therefore, cannot account for radio
observations of SNe, either.

Other parametrizations of the mass loss rate in the HR diagram have
been proposed by different authors (\eg De Jager, Nieuwenhuijzen \& van
der Hucht 1988, Salasnich, Bressan and Chiosi 1999), but insofar for
RSGs the main dependence of $\dot{M}$ is a power of $\sim 2$ of the
luminosity, they all are unable to reproduce appreciable  mass loss
variations over a timescale of roughly 10$^4$ years.

Actually, one notices that for masses above 10 \msun, the last phases
of the RGS evolution fall within the extrapolation of the Cepheid
instability strip (see Figure 4), as calculated by Bono \etal~ (1996),
and therefore, one may expect that pulsational instabilities could
represent the additional mechanism needed to trigger high mass loss
rates.  Indeed,   the pioneering work of Heber \etal~ (1997), based on 
both  linear and nonlinear pulsation models, demonstrated  that RSG
stars are pulsationally  unstable.  In particular, they found that, for
periods approaching the  Kelvin-Helmotz time scale, these stars display
large luminosity amplitudes,  which could trigger a strong enhancement
in their mass loss rate before they  explode as supernovae. According
to these authors this pulsation behaviour  should take place during the
last few $10^4$ yrs before the core collapse,  due to the large
increase in the luminosity to mass ratio experienced by RSG  stars
during these evolutionary phases.

However, the nonlinear calculations performed by Heber \etal~ (1997) 
were hampered by the  fact that their hydrodynamic code  could not
properly handle  pulsation destabilizations characterized both by small
growth rates due to  numerical damping, and by large pulsation
amplitudes due to the formation  and propagation of strong shock waves
during the approach to limit cycle  stability. Also, as Heber \etal~
(1997) pointed out, their main theoretical  difficulty in dealing
with the dynamical instabilities of RSG variables resided in   the
coupling between convection and pulsation. In fact, they constructed
the linear models by assuming that the convective flux is  frozen in,
and  the nonlinear ones by assuming that the convective flux  is
instantaneously  adjusted. However,  this treatment does not  account
for the driving and/or quenching  effects caused by the interaction
between pulsation and convection: this shortcoming  may explain why their
nonlinear models could not approach a stable  limit cycle.

It is clear that a more general approach  must be adopted to solve the
problem.  This motivated us to start a systematic study of the
pulsational properties of massive stars.  In the following we shall
illustrate briefly the procedures adopted and the first results
obtained (Section 2), and will present and discuss our findings on the
mass loss rates in  the late phases of the evolution of massive stars
(Sections 3 and 4). 

\begin{figure}
\centerline{
\psfig{figure=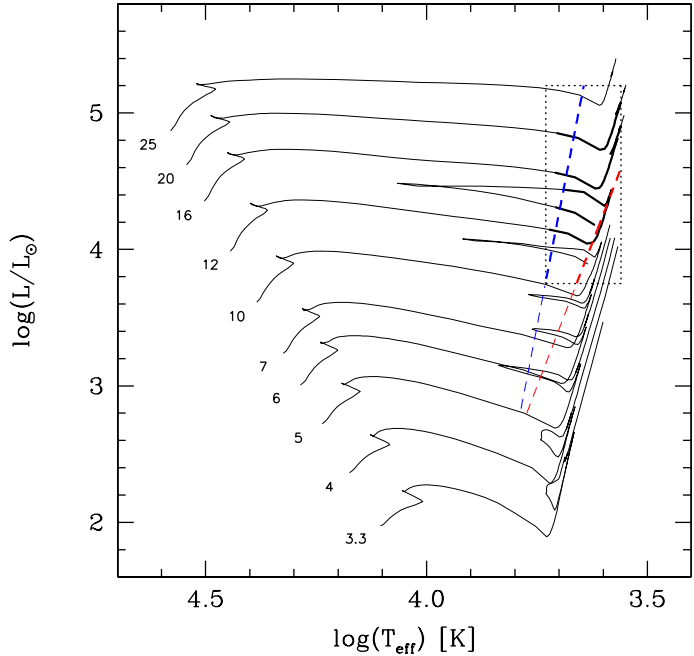,width=10cm}}  \caption{Evolutionary tracks
for stars in the range 3.3--25 \msun. The marked area is the portion of
the HR diagram shown in more detail in Figure 5. The dashed lines are
the extrapolation  of the fundamental boundaries of the Cepheid
instability strip  according to the analytical relations provided by
Bono \etal~(2000a), which are based on a detailed  investigation of
Cepheid models at solar chemical composition and stellar  masses
ranging from 5 to 11 \msun. Note that a substantial portion of both
H-shell and He burning phases for stars with masses higher than about
10 \msun~ occur within the instability strip.}
\label{hrdiag99}
\end{figure}

\begin{figure}
\centerline{
\psfig{figure=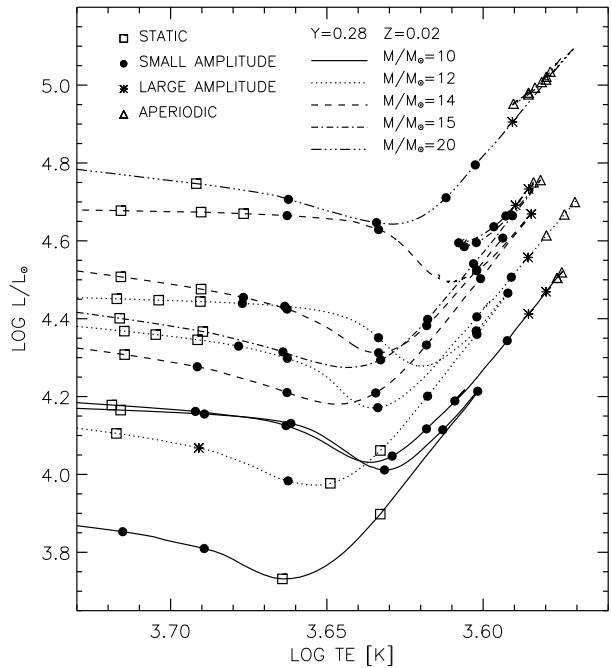,width=9cm}}
\caption{Evolutionary tracks and pulsation models at solar chemical 
composition in the HR diagram. Evolutionary models constructed by 
adopting different stellar masses are plotted by using different line
styles,  while pulsation models characterized by different limiting
amplitude behaviours   are plotted using different symbols. See text for
further details.}
\label{rosse}
\end{figure}

\section{Theoretical Framework} 

The procedures employed to construct both linear and nonlinear  models
of high-mass radial variables have been described in detail in a number
of papers  (Bono \& Stellingwerf 1994; Bono, Caputo, \& Marconi 1998;
Bono, Marconi \&  Stellingwerf 1999), so that a  brief outline of the
methods adopted will suffice here.  In particular: \\
$\bullet$ We constructed a set of limiting amplitude, nonlinear, convective
models  of Red Super Giant (RSG) variables during both hydrogen and helium
burning  evolutionary phases. \\
$\bullet$ A Lagrangian one-dimensional hydrocode was used in which local
conservation equations are simultaneously  solved with a nonlocal,
time-dependent convective transport equation (Stellingwerf 1982; Bono \&
Stellingwerf 1994; Bono et al. 1998). \\
$\bullet$ Nonlinear effects such as the coupling  between convection and
pulsation, the convective overshooting and the  superadiabatic gradients are
fully taken into account. \\
$\bullet$ As for the opacity, which is a key ingredient for 
constructing stellar envelope models, we adopted OPAL opacities
(Iglesias \& Rogers 1996) for T $>$ 10,000 K  and molecular opacities
(Alexander \& Ferguson 1994) T $<$ 10,000K.  The method adopted for
handling opacity  tables was discussed in Bono et al. (1996). \\
$\bullet$ Artificial viscosity was included following Stellingwerf (1975) 
prescriptions.\\
$\bullet$ To provide accurate predictions on the limit cycle behaviour 
of these objects, the governing equations were integrated in time for a 
number of periods ranging from 1000 to 5000.  

\begin{figure}
\centerline{
\psfig{figure=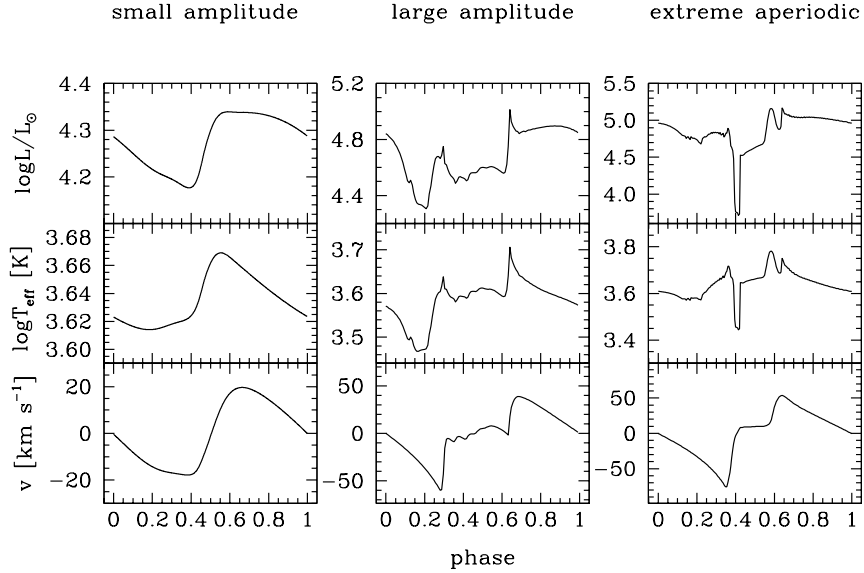,width=12cm,angle=-90}}
\caption{The cycle variation of luminosity, temperature and velocity
for a 15 $M_\odot$ red supergiant for three distinct regimes, small
amplitude (left panel), large amplitude (central panel), and aperiodic
extreme pulsations (right panel).}
\label{pulsation}
\end{figure}

In order to make a detailed analysis of RSG pulsation properties during
both H and He burning phases, to be compared with actual properties  of
real RGS stars, we constructed several sequences of models at fixed
chemical composition (Y=0.28, Z=0.02) which cover a wide range of
stellar masses $10 \le$ \msun $\le 20$. Moreover, since we are
interested in mapping the properties of RSG stars from H-shell  burning
up to the central He exhaustion, both the luminosities and the
effective temperature values were selected directly along the
evolutionary tracks (see Figure 4). The evolutionary calculations were
performed at fixed mass -- \ie no mass-loss -- and neglecting the
effects of both convective core overshooting and rotation. Since the
pulsational properties of a star depend mostly on the physical
structure of the envelope regions in which H and He  undergo partial
ionization, \ie well above the layers in which the  nuclear burning
takes place, the use of constant mass evolutionary tracks for our
study  does not limit the qualitative value of our conclusions.
However,  self-consistent evolutionary  models which properly include a
parametrization of mass loss will eventually be needed for a full,
quantitative description of the phenomenon (see Section 4).

The input physics and physical assumptions adopted for  constructing
evolutionary and pulsational models will be described in  detail in a
forthcoming paper (Bono \& Panagia 2000).   Figure 5 shows the location
in the HR diagram of both evolutionary tracks  and pulsation models we
constructed. The pulsation models characterized  by different limiting
amplitude behaviour are plotted with different  symbols. The squares
refer to models which are pulsationally stable, \ie those in which,
after the initial perturbation, the radial motions decay and the
structure  approaches once again the static configuration. Filled
circles and  asterisks refer to models which show small ($\Delta
logL<0.4$)and large ($\Delta logL>0.4$)  pulsation amplitudes together
with a periodic behaviour (stable limit cycle).  Triangles denote the
models which not only present large pulsation amplitudes but also
aperiodic radial displacements (unstable limit cycle). The behaviour of 
pulsation properties discloses several interesting features:\\ 
 1) For effective temperatures lower than approximately 5100 K
high-mass  models are, with few exceptions, pulsationally unstable in
the  fundamental mode both during H shell and He burning phases. \\  
 2) The pulsational behaviour is mainly governed by the effective
temperature and to a less extent by the luminosity. In fact, the
transition from small  to large pulsation amplitudes ($T_e \approx3900$
K) and from periodic to  aperiodic behaviour ($T_e \le 3800$ K) take
place roughly at constant  temperature.  \\
 3) Current models support the evidence suggested by 
Li \& Gong (1994) on the basis of linear, nonadiabatic models that 
RSG variables are pulsating in the fundamental mode. In fact we find 
that throughout this region of the HR diagram the fundamental mode 
is pulsationally unstable, whereas the first overtone is stable.\\
 4) The region in which the models attain small amplitudes is the
natural  extension of the classical Cepheid instability strip. This
confirms the empirical evidence originally brought out by Eichendorf
\& Reipurth (1979) and more recently by Kienzle et al. (1998), as well
as the theoretical  prediction by Soukup \& Cox (1996). \\ 
5) Interestingly enough, theoretical light curves of periodic 
large amplitude models show the characteristic RV Tauri behaviour, i.e. 
alternating deep and shallow minima, observed in some RSG variables 
(Eichendorf \& Reipurth (1979).

\section{Pulsationally Induced Mass Loss}

As discussed above, in the reddest part of their RSG evolution, massive
stars are found to pulsate with large amplitudes in both luminosity and
velocity. In these phases, the radial velocity near the stellar surface
may reach values of 50 \kms~ or higher, which may become higher than
the effective escape velocity, \ie the one computed by including both
inward gravitational force and outward radiation acceleration. 
Therefore, the outer layers may become unbound and be lost from the
system,  thus producing a high mass loss rate. As an illustration,
Figure 6 shows the variation of luminosity, temperature and velocity
for a 15 $M_\odot$ red supergiant for three distinct regimes, small
amplitude (left panel), large amplitude (central panel), and aperiodic
extreme pulsations (right panel).

\begin{figure}
\centerline{
\psfig{figure=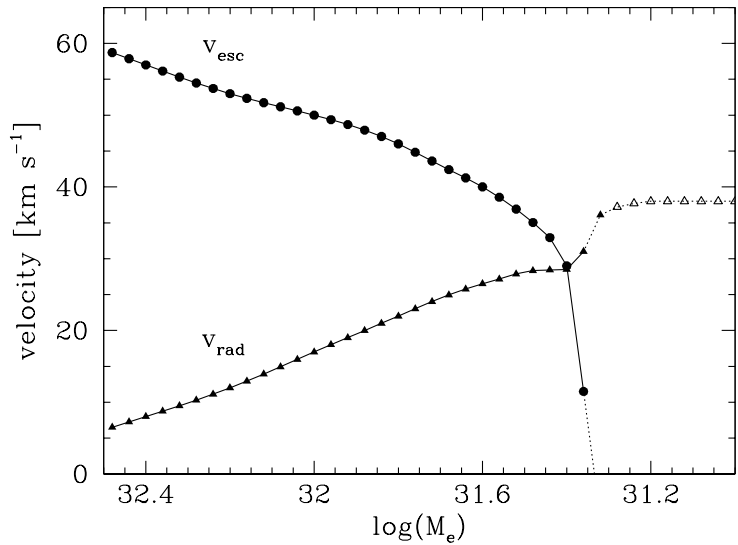,width=11cm,angle=-90}}
\caption{Radial velocity and escape velocity as a function of the mass
as measured from the stellar surface for a 15 $M_\odot$ star at
T$_{eff}$=3836~K and log(L/L$_\odot$)=4.75.}
\label{vel_mass}
\end{figure}

\begin{figure}
\centerline{
\psfig{figure=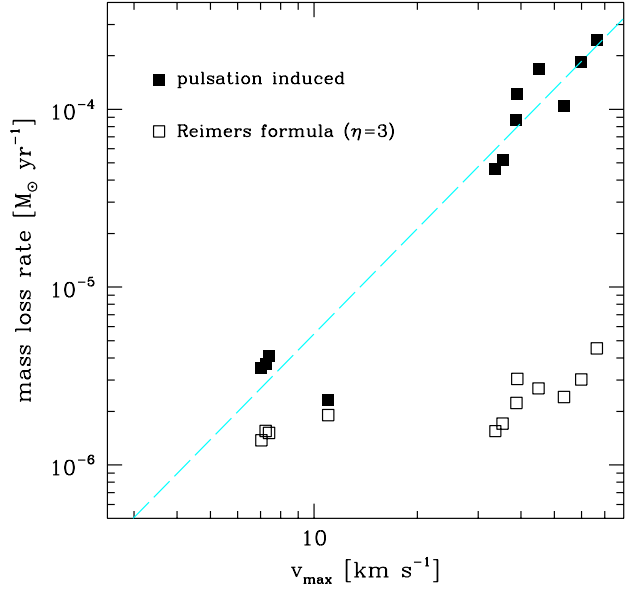,width=9cm}} \caption{Mass loss rates as a
function of  the maximum velocity (see text). Filled symbols represent
the pulsation-induced mass loss rates, and the open symbols the rates
computed with Reimers' formula. The dashes line is a linear best-fit to
the pulsation induced mass loss rates.}
\label{calib}
\end{figure}

In order to calculate pulsation induced mass loss, for each model we 
identify the outer layers for which 

$$  v_{rad} ~ > ~v_{esc} + c_s	$$

\noindent where $v_{rad}$ is the radial velocity of a given layer,
$v_{esc}$ is the effective escape velocity  that includes the effects
of radiation forces, and  $c_{s}$ is the sound speed (Hill \& Willson
1979). Those layers are effectively unbound and, therefore, are lost
from the star. An example is shown in Figure 7 where we plot the actual
velocity of the stellar envelope layers as a function of the stellar
radius and compare them with the effective escape velocity.  The mass
present above the radius where the the actual velocity exceeds the
escape velocity reprents the amount of mass which is lost in a
pseudo-impulsive event. 

The characteristic time between successive pseudo-impulsive events is
the Kelvin-Helmotz time, \ie 

$$ \tau_{KH} ~ \simeq~ \frac{GM^2}{RL} = 63~ [\frac{M}{10M_\odot}]^2
~[\frac{R}{500R_\odot}]^{-1}   ~[\frac{L}{10^5L_\odot}]^{-1}	~ yrs$$ 

\vskip.1in

Thus, the mass loss rate is given by 

$$ \dot{M} ~= ~\frac{M(v>v_{esc}+c_s)}{\tau_{KH}}  $$

An inspection to Figure 7 shows that layers as massive as 10$^{-2}$
\msun may become unbound and be lost from the stellar surface within
time intervals of several tens of years, thus producing mass loss rates
of the order of several 10$^{-4}$ \mloss or even higher.

Following this recipe, we have determined the mass loss rates for
several values of the stellar mass and for a variety of  locations in
the HR diagram. As shown in Figure 8,  we find that the pulsation
induced mass loss can satisfactorily be represented as a power law
function of the maximum expansion pulsational velocity (\ie the maximum
velocity that is attained by the outermost layers in a pulsational
cycle), \ie 

$$ log(\dot{M})~ =~ -7.24~ +~ 1.97 \times log(v_{max})	$$

Since pulsation-induced mass loss is an additional mass loss mechanism
that comes on top of more conventional radiation-pressure induced mass
loss, the total mass loss rate is assumed to be the straight sum of the
two rates. 

\begin{figure}
\centerline{
\psfig{figure=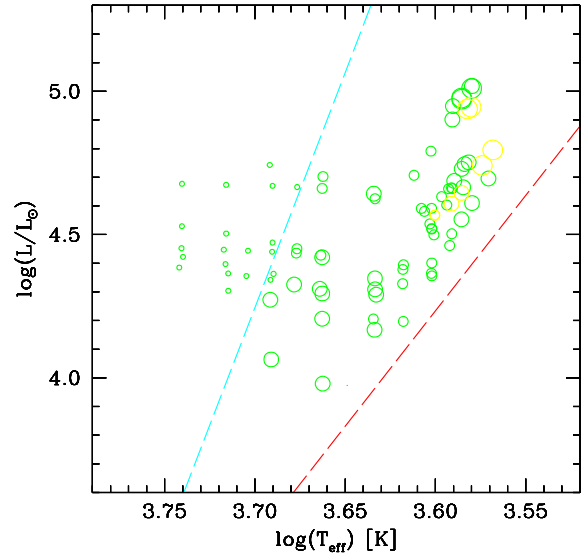,width=8.5cm}}
\caption{Pulsationally enhanced mass loss rates as a function of the
position in the HR diagram. The size of the circles are proportional to
the logarithm of the mass loss rate up to a maximum value of almost
10$^{-3}$ \mloss. Note the strong enhancement of the mass loss rate
within  the Cepheid instability strip (area within the dashed lines;
Bono \etal~ 2000a).}
\label{mlosses} 
\end{figure}
\begin{figure}
\centerline{
\psfig{figure=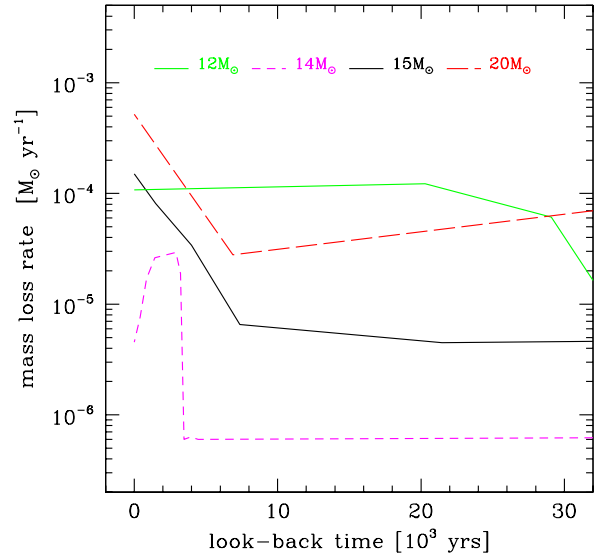,width=8.5cm}} 
\caption{Pulsationally enhanced mass loss rates as a function 
of look-back time (short time-scale variation).}
\label{doeswork1}
\end{figure}

The main result is that that including the effects of  pulsations, the
predicted mass loss rate in the RGS phase is a strong function of both 
the luminosity and the temperature, in the sense that the mass loss 
process is strongly enhanced by pulsations when a star is moving toward
cooler effective temperatures. Figure 9 illustrates this  result and
shows that now it should only take small adjustment to fully  reproduce
the observations. Moreover,  since a  red supergiant may cross  the 
instability  strip rather quickly, its mass loss can change
considerably  on relatively short time scales, thus accounting for
another observational fact.   It is worth mentioning that Feast (1991)
found, on the basis of IRAS  data for 16 RSG variables in the LMC, a
period-mass loss relation. This  relation supports a similar behaviour
i.e. a steady increase in the  mass-loss when moving from short to long
period variables.

\section{Discussion and Conclusions}

The computed mass loss rates, for stars in the range 12-20 M$_\odot$,
as a function of look-back time  are displayed in Figures 10-12 for
short (0-30,000 years), medium (0-150,000 years)  and long time-scales
(0-1 Myrs), respectively. We see that the mass loss rates may be as
high as almost 10$^{-3}$ \mloss, \ie similar to what is measured for
extreme  red supergiants, and may vary by an order of magnitude over
relatively short times, say, 10,000 years or less.  In other words the
predicted mass loss rates are able to account, at least qualitatively,
for all of the features observed in radio supernovae. Moreover,  since
the predicted mass  loss history is a critical function of how a
massive star evolves within  the pulsation instability strip, a
comparison between observations and  theory should lead to an accurate
determination of the stellar  progenitor mass. For example, the mass
loss decline of a 20 \msun star may be used to represent the apparent
drop of emission of SN~1988Z  about 9 years after explosion (cf. Figure
1).  Similarly, the quick increase found for our 14~\msun model closely
resembles the behaviour observed for SN~1993J.  Of course, detailed
comparisons will be meaningful only we will have a fully
self-consistent set of evolutionary tracks (see below).

The mid- and long-term behaviour of the mass loss rate as a function 
of look-back time is also interesting because allows one to make
predictions about the radio emission, as well as on {\it any} other
phenomenon linked to a SN shock front and/or ejecta interaction with a
dense circumstellar medium, such as relatively narrow optical emission
lines and X-ray emission. As we can see in Figures 11 and 12, massive
stars are expected to display rather sudden variations of  their mass
loss rates of all time scales, both because of pulsational
instabilities which arise with crossing the instability strip (\eg the
12~\msun star in the time range 20-60$\times 10^3$ years) and because
of the so-called blue loops (an effect clearly apparent at look-back
times around 0.4--1 Myrs) that are determined by a combination of core
He-burning and shell H-burning (\eg Brocato \& Castellani 1993, Langer
\& Maeder 1995).  Because of these effects, one may expect that in some
cases, a SN may drop below detection limit for a while but still may
have a renaissance, in the X-ray, optical and radio domains,  several
tens or hundreds of years later.

\begin{figure}
\centerline{
\psfig{figure=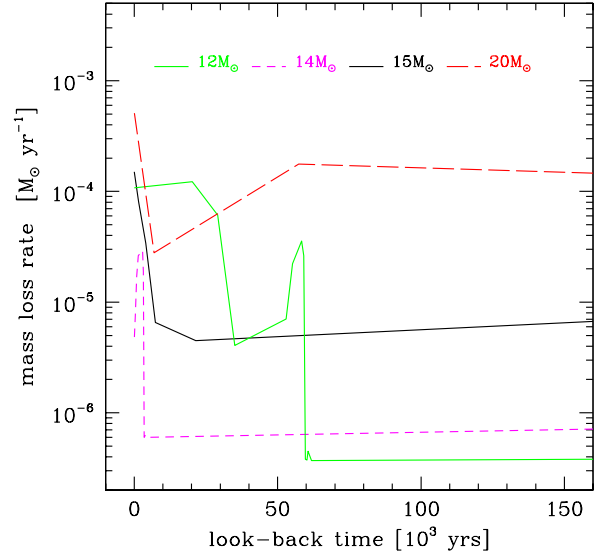,width=8.5cm}} 
\caption{Pulsationally enhanced mass loss rates as a function 
of look-back time (medium time-scale variation).}
\label{doeswork2}
\end{figure}
\begin{figure}
\centerline{
\psfig{figure=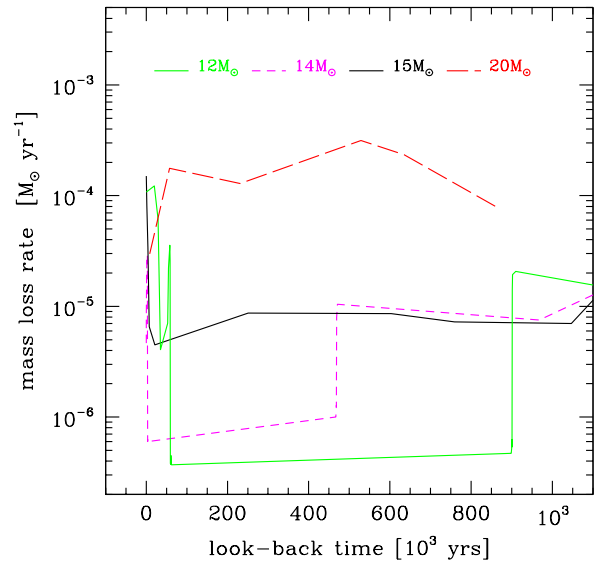,width=8.5cm}} 
\caption{Pulsationally enhanced mass loss rates as a function 
of look-back time (long time-scale variation).}
\label{doeswork3}
\end{figure}

Also, we note in passing that our findings support the  empirical
evidence recently brought out by van Loon et al. (1999) on  the basis
of ISO data on RSG stars in the LMC. In fact, they found  that the mass
loss rates increase with increasing luminosities  and decreasing
effective temperatures and range from $10^{-6}$ up  to $10^{-3} M_\odot
yr^{-1}$. A strong dependence of the mass loss  rate on the effective
temperature in Tip-AGB stars was recently  suggested by Schr\"oder,
Winters and Sedlmayer (1999) on the basis  of theoretical evolutionary
models which account for carbon-rich  wind driven by radiation pressure
on dust.

Another interesting consequence of our results is that  a more
efficient mass loss in the RSG phase implies a lower mass cutoff to
produce Wolf-Rayet stars and, therefore, one has to expect  a more
efficient mass return into the ISM than commonly  adopted in galactic
evolution calculations.

Still there are improvements and refinements to apply to our models,
because the calculations we presented here are not fully
self-consistent in that we adopted evolutionary tracks computed either
with no mass loss whatsoever, or with modest mass loss rates, and on
them we performed our pulsational stability analysis and, thus,
determined our new mass loss rates.  Moreover, our models were
constructed by adopting the diffusion  approximation even in optically
thin layers and therefore we neglected  the dust formation processes
(Arndt et al. 1997). A macroscopic example of the shortcomings of our
current approach is that if we integrate the mass loss rates over time,
in many cases we find that the star looses a substantial fraction of
their mass before reaching its evolutionary end.  Although this is
close to what one should expect on the basis of observations,
definitely it is at variance with the assumptions that went into the
adopted evolutionary model calculations.

 It is clear that what we need to do now is to follow an iterative
procedure in which we first use our present prescriptions to compute
new evolutionary tracks, then we repeat our pulsational stability
analysis, then we compute new mass loss rates, and we iterate the
procedure until adequate convergence is achieved. This work is in
progress and will be presented in future papers.
For the time being, our conclusions can be summarized as follow: \\
-- We have defined a new theoretical scenario for pulsation induced 
mass loss in RSGs.\\
-- RSGs are pulsationally unstable for a substantial portion of their
lifetimes.\\
-- Dynamical instabilities play a key role in driving mass loss.\\
-- Bright, cool RGSs undergo  mass loss at considerably higher rates 
than commonly adopted in stellar evolution.\\
-- Comparisons of model predictions with observed CSM phenomena around
SNII will provide valuable diagnostics about their progenitors and
their evolutionary history.\\
-- More efficient mass loss in the RSG phase implies a lower mass
cutoff to produce Wolf-Rayet stars and a more efficient return of
polluted material into the ISM, thus affecting the expected chemical
evolution of galaxies.


\end{document}